\documentclass[twocolumn,showpacs,preprintnumbers,amsmath,amssymb]{revtex4}
\usepackage{graphicx}
\usepackage{dcolumn}
\usepackage{bm}
\begin{document}
\title{Statistical analysis of a dynamical multifragmentation path}
\author{A. H. Raduta$^{a,c}$, M. Colonna$^{a,b}$, V. Baran$^{a,c}$, 
M. Di Toro$^{a,b}$}
\affiliation{
        $^a$LNS-INFN, I-95123, Catania, Italy\\
	$^b$Physics and Astronomy Dept. University of Catania, Italy\\
        $^c$NIPNE, RO-76900 Bucharest, Romania}

\begin{abstract}
A microcanonical multifragmentation model (MMM) is used for 
investigating whether
equilibration {\em really} occurs in the dynamical evolution of two heavy ion collisions simulated
via a stochastic mean field approach (SMF). The standard deviation function between the {\em
dynamically} obtained freeze-out fragment distributions corresponding to the reaction 
$^{129}$Xe+$^{119}$Sn at 32 MeV/u and the MMM ones corresponding to a wide 
range of mass, excitation energy, freeze-out volume and nuclear level density cut-off parameter 
shows a unique minimum. A distinct statistically equilibrated stage is identified in the dynamical
evolution of the system.
\end{abstract}
\maketitle

It is known from more than 15 years that violent heavy ion collisions
lead to an advanced disassembly of the compound system known under the
name of nuclear fragmentation.
Good agreements between various
observables related to the asymptotically resulted fragments and various models assuming
statistical equilibrium 
\cite{Gross1,Bondorf,Randrup,mmm}
lead to the conclusion
that a huge part of the available phase space is populated during the 
fragmentation process. Hence the break-up of a statistically equilibrated nuclear source
could be at the origin of fragment production.  
The source size, its excitation energy 
and its volume are thus quantities which can only be {\em indirectly}
evaluated by comparisons between experimental data and 
statistical multifragmentation model predictions via a back-tracing 
procedure. 
However, the comparison
process is complicated by the presence of several effects, 
such as pre-equilibrium particle emission,
collective radial expansion (see e.g. \cite{philippe}), Coulomb propagation of the break-up primary 
fragments, secondary particle emissions. 
So it is difficult to ascertain whether these quantities really correspond
to the source properties obtained at a given time during the fragmentation path.
Moreover, there are intrinsic dynamical
characteristics such as the freeze-out specific time, directly related to 
parameters such as the level density cut-off parameter, $\tau$ (see Ref. \cite{prc2002}) contributing to 
the weights of the system statistical ensemble. In the absence of any direct information 
about the freeze-out this last parameter has to be 
{\em employed as a fitting  parameter} in a statistical model. In Ref. \cite{prc2002}, a very good 
agreement between statistical fragment distribution predictions and experimental data was obtained 
{\em assuming} $\tau=9$ MeV. However, one has to have {\em direct} access to the freeze-out
events in order to unambiguously decide on the value of such parameters and, if an equilibrated 
source exists, to find its location in space and time. This task can be achieved using ``freeze-out'' 
information from a dynamical model.

This letter investigates whether statistical equilibration occurs in the dynamical 
path of two heavy ion collision and, if so, which are the corresponding freeze-out parameters. To this aim, 
the ``freeze-out'' data of a stochastic mean field (SMF)
approach \cite{buu} is analyzed via a sharp microcanonical
 multifragmentation model (MMM) \cite{mmm,prc2002}. 

We use the stochastic mean-field approach introduced in ref.\cite{mac}.
According to this theory, the fragmentation process
is dominated by the growth of volume (spinodal) and surface instabilities encountered during the expansion phase 
of the considered excited systems. 
The dynamical evolution of the system is still described in terms of the one-body distribution 
function (mean-field description), 
however this function experiences a stochastic evolution, in response to the action of the 
fluctuation term. 
The amplitude of the stochastic term incorporated in the treatment is essentially
determined by the degree of thermal agitation present in the system. Hence fluctuations provide the seeds of the
formation of fragments, whose characteristics are related to the properties of the most unstable
collective modes \cite{rep}.

In the model \cite{mac} fluctuations are implemented only in {\bf r} space. Within the assumption
of local thermal equilibrium, the kinetic fluctuations typical of a Fermi gas are projected on density fluctuations.
Then fluctuations are propagated by the unstable mean-field, leading to the disassembly of the system.  

The MMM model concerns the disassembly of a statistically equilibrated nuclear source $(A,Z,E,V)$ (i.e. the 
source is defined by the parameters: mass number, atomic number, excitation energy and freeze-out volume 
respectively). Its basic hypothesis is equal probability between all configurations $C:\{A_i,Z_i,\epsilon_i, 
{\bf r}_i,{\bf p}_i,~~i=1,\dots,N\}$ (the mass number, the atomic number, the excitation energy, the position
and the momentum of each fragment $i$ of the configuration $C$, composed of $N$ fragments) which are subject
to standard microcanonical constraints: $\sum_i A_i=A$, $\sum_i Z_i=Z$, $\sum_i {\bf p}_i=0$, $\sum_i {\bf 
r}_i\times{\bf p}_i=0$, $E$ - constant. The fragment level density (entering the statistical weight of a 
configuration) is of Fermi-gas type adjusted with the cut-off factor $\exp(-\epsilon/\tau)$: 
$\rho(\epsilon)=\rho_0(\epsilon) \exp(-\epsilon/\tau)$ \cite{prc2002}. 
The above factor counts for the dramatic decrease of 
the lifetime of fragment excited states respective to the freeze-out specific time as the excitation energy 
increases (i.e. earlier freeze-outs should correspond to larger values of $\tau$). The model can work within two freeze-out hypotheses: (1) fragments are treated as hard spheres placed into a spherical freeze-out recipient and are not allowed to overlap each-other or the recipient wall; (2) fragments may be deformed and a corresponding free-volume expression is approaching the integration over fragment positions \cite{prc2002}. Though more schematic, hypothesis (2) seems more adequate for the present study, so we use it herein. Further a Metropolis-type simulation is employed for determining the average value of any system observable (see Refs. \cite{mmm} for more details). While the model includes secondary deexcitation, this stage is not needed for this investigation so we will use here only its {\em primary decay} stage.

Using MMM we investigate whether the ``primary events'' produced by the stochastic mean-field approach as a 
result of $^{129}$Xe+$^{119}$Sn at 32 MeV/u reaction may correspond to the statistical equilibration of 
the compound system.  
We consider only very central collisions of the above reaction. According to the dynamical simulations
performed in Ref.\cite{buu}, it is observed that, after the initial collisional shock, the system expands
towards low densities entering the unstable region of the nuclear matter phase diagram (after about
$100~fm/c$ from the beginning of the reaction). 
Then fragments are formed through spinodal decomposition.  
The {\em dynamical freeze-out} time is defined as the 
time when the fragment formation process is over. Hence average fragment multiplicities and distributions do
not evolve anymore. For the reaction considered this time is $240~fm/c$.

Our aim is to investigate whether fragment distributions are compatible with the statistical
phase space occupancy, as predicted by MMM.    
For washing-up pre-equilibrium effects which should appear in the dynamical simulation, 
only intermediate mass fragments (IMF) (i.e. fragments with $Z\ge3$) are selected. Therefore, all comparisons
 between MMM and stochastic mean-field results are to be restricted to IMF's. 
Due to the large Coulomb repulsion  among primary fragments it is reasonable to assume that the largest uncertainty 
in the equilibrated source estimated from the dynamical approach concerns the volume.
In other words while {\em real} equilibration may occur at a different volume, than the one corresponding to the ``dynamical'' data, we assume that fragment sizes and excitation energies are roughly preserved. 
Therefore, we will fit
the fragment size distributions and their internal excitation energy but {\em not} the volume. 
The best fit can be found by minimizing the following error function:
\begin{eqnarray}
{\cal E}&=&\{3\left[f(\left<A_{bound}\right>)+f(\left<Z_{bound}\right>)\right] \nonumber \\
        &+&\left[\sum_{N_{IMF}}f[\left<{\rm d}N/{\rm d}N_{IMF}\right>]/\sum_{N_{IMF}}1\right] \nonumber \\
	&+&f(\left<\epsilon_{IMF}\right>) 
        +\sum_{i=1}^3 f(\left<Z_{{\max}i}\right>)/3\}/9
\end{eqnarray}
where $\left< \cdot \right>$ stands for average, $A_{bound}$ and $Z_{bound}$ are the bound mass and 
charge (sum of the mass number and, respectively, atomic number of all IMF's from a given event), $N_{IMF}$ is
the number of IMF's, $\epsilon_{IMF}$ is the fragment excitation energy per nucleon and $Z_{{\max}i}$ with $i=1,2,3$ are 
the largest, second largest and third largest charge from one fragmentation event. Further, 
$f(x)=\left|2(x_s-x_d)/(x_s+x_d)\right|$, where the indexes $s$ and $d$ stand for ``statistic'' and
``dynamic'' and $\left|x\right|$ is the absolute value of $x$. One may observe in eq. (1) that the bound mass
 and charge terms are overweighted with an (arbitrary) factor $3$. Of course, this does not influence the
  significance
of the error function, $\cal E$, which will vanish for a perfect agreement between the $s$ and $d$ results. 
On the other side, this will put stronger constraints on source size identification. 
\begin{figure}
\includegraphics[height=12.5cm]{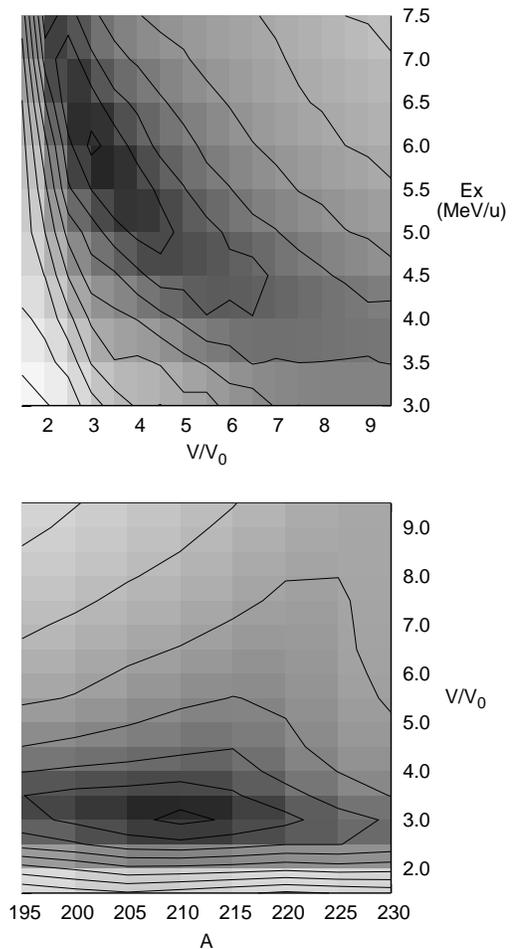}
\caption{Contour plots of the error function $\cal E$ [see eq. (1)]: in the $(V/V_0,E)$ plane corresponding to $A=210$ (upper panel); in the $(A,V/V_0)$ plane corresponding to $E=5.7$ MeV/u (lower panel). Darker regions correspond to smaller $\cal E$; units are relative. \\[-.7cm]}
\end{figure}
For finding the best agreement between 
statistical and dynamical results we variate the MMM parameters $A$, $E$, $V$ and $\tau$ in {\em
 sufficiently} wide 
ranges thus constructing a four-dimensional grid. The ranges are $A:\left[195,230\right]$,
 $E:\left[3,7.5\right]$
 MeV/u, $V/V_0:\left[1.5,9.5\right]$, $\tau=12, 16, {\infty}$ MeV. The source is considered to have the $A/Z$ ratio of the $^{129}$Xe+$^{119}$Sn reaction. Considerable computing power was involved
 for obtaining this result in reasonable time.
 An absolute minimum of the error function, $\cal E$, was found at $A=210$, $Z=87$, $V/V_0=3.4$, $E=5.7$
 MeV/u, $\tau=\infty$. Cuts in $\cal E$ corresponding to $A=210$ and $E=5.7$ MeV/u are represented in the 
 upper part and respectively the lower part of Fig. 1. The evolution of $\cal E_{\min}$ (i.e. the minimum
 $\cal E$ corresponding to a given $\tau$) with $\tau$ is represented in Table \ref{tab:1}. 
 From the above mentioned 
 figure one can see that the global minimum is clearly determined and there are no secondary minima. 
\begin{table}
\caption{\label{tab:1} Evolution of $\cal E_{\min}$ (see text) with $\tau$.}
\begin{ruledtabular}
\begin{tabular}{rccc}
$\tau$ & 12 & 16 & ${\infty}$ \\
$\cal E_{\min}$ & 0.15 & 0.11 &0.06\\
\end{tabular}
\end{ruledtabular}
\end{table}

The statistical source corresponding to the minimum value of $\cal E$ yields the following results:
$\left<A_{bound}\right>=199.03$, $\left<Z_{bound}\right>=83.74$, $\left<\epsilon_{IMF}\right>=4.21$ MeV/u,
$\left<Z_{\max1}\right>=42.34$, $\left<Z_{\max2}\right>=24.35$, $\left<Z_{\max3}\right>=11.4$. These are to 
be compared with the corresponding {\em dynamical} results: $\left<A_{bound}\right>=199$, 
$\left<Z_{bound}\right>=84$, $\left<\epsilon_{IMF}\right>=4.3$ MeV/u, $\left<Z_{\max1}\right>=41.95$, 
$\left<Z_{\max2}\right>=22.5$, $\left<Z_{\max3}\right>=13.3$. 
Note the excellent agreement for all considered 
observables, proving the very good quality of the fit.   
That means that the fragment size related 
features obtained dynamically have been reproduced by the primary decay stage of the MMM model. 
This can be 
seen in Fig. 3, as well, where the {\em dynamical} $Z$ and $N_{IMF}$ distributions fit perfectly the MMM ones. 
\begin{figure}
\includegraphics[height=14cm]{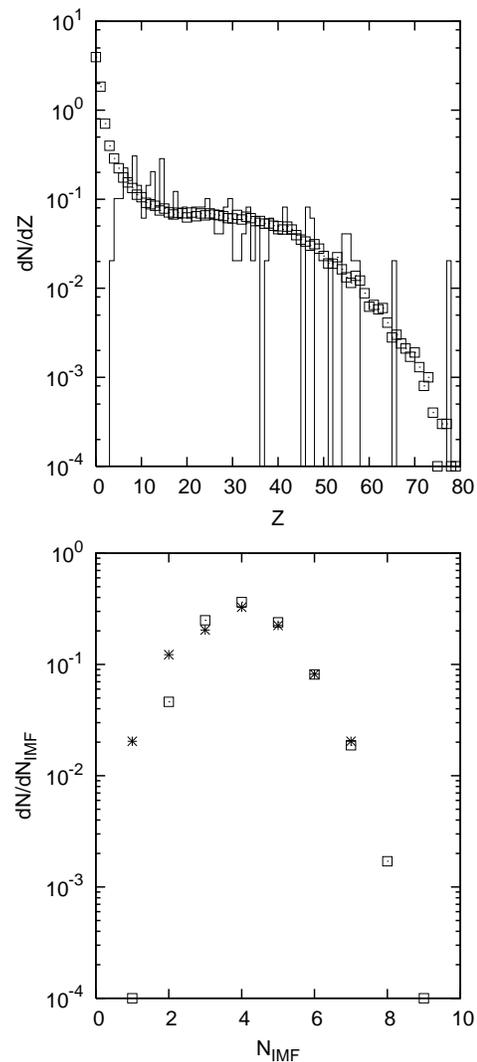}
\caption{Dynamical charge (upper panel) and number of IMF distributions (lower panel) in comparisons with the statistical ones. Statistical results are represented by open squares; histogram (upper panel) and stars (lower panel) corespond to the dynamical ones.\\[-.7cm]}
\end{figure}
\begin{figure}
\includegraphics[height=14cm]{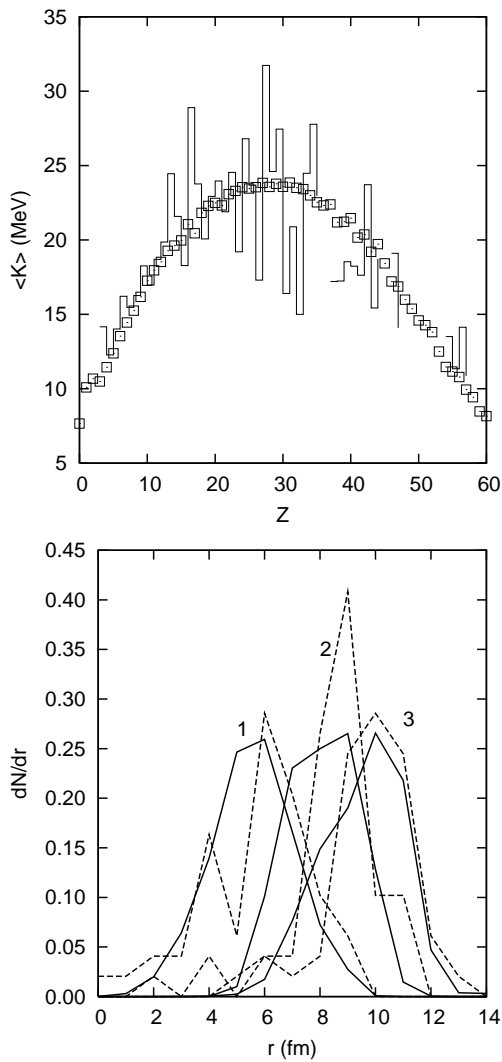}
\caption{Upper panel: ``statistical'' (open squares) and ``dynamical'' (histogram) fragment average 
kinetic energy versus charge. Lower panel: ``statistical'' (solid lines) and ``dynamical'' (dashed lines) radial distribution of fragments with largest (peak ``1''), second largest (peak ``2'') and third largest (peak ``3'') charge in one fragmentation event. The plot corresponds to $\left<(\widetilde{V}/V_0)_{IMF}\right>=9.08$.\\[-.7cm]}
\end{figure}
Due to the fact that the fragment size distributions and fragment excitations were used for constructing
$\cal E$ these results are natural, given the {\em small} value of the obtained minimum. At this stage the
question still remains: {\em even if fragment size distributions and excitation energies of the  fragments are very well reproduced, do the dynamically formed fragments come from an equilibrated source with freeze-out volume $V=3.4V_0$, as predicted by MMM?}

In the present work the freeze-out volume $V$ is the volume of the smallest sphere which {\em totally}
includes all fragments. We denote by $\widetilde{V}$ the volume of the smallest sphere which {\em totally} 
includes all  fragments {\em and} has the center located in the center of mass of the system. Obviously, 
$\widetilde{V}>V$. The ``dynamical'' events have $\left<(\widetilde{V}/V_0)_{IMF}\right>=9.08$; the 
statistical ones have $\left<(\widetilde{V}/V_0)_{IMF}\right>=4.93$ (the $IMF$ index indicates that we refer
 to the volume occupied by IMF fragments). 
This means that equilibration, as predicted by MMM, may have occured at an earlier time compared to the considered dynamical events, i.e. pre-fragments already appear at an earlier time and they are actually equilibrated inside a
smaller volume.

One can simply test this hypothesis: one just has to propagate the fragments in their mutual Coulomb field from the 
freeze-out positioning as generated by MMM up to $(\widetilde{V}/V_0)_{IMF}=9.08$ (i.e. the value 
corresponding to the ``dynamical'' events) and then compare ``dynamical'' and ``statistical'' fragment 
kinetic energies and positions. 
However, in performing such a comparison, one has to guess the initial fragment velocities, 
that could have a collective component. Flow is easily accontable for in MMM \cite{prc2002}. 
The best reproduction of the dynamical results 
is obtained for a flow energy equal to zero. (Note that this result is in agreement with the dynamical simulations where
the initial expansion energy present at early times is reduced by the strong interaction between prefragments and finally flow becomes close to zero.) The time of the Coulomb propagation is around 100 fm/c.
The comparison is presented in Fig. 4: the fragment average kinetic energies 
 versus charge is represented in the upper panel; the radial distribution of the fragment with 
 largest, second largest and third largest charge is represented in the lower panel. 
An excellent agreement is observed between ``dynamical'' and ``statistical'' data for both observables indicating the physical consistency of the scenario used in the present {\em statistical} analysis. {\em It means that the dynamically generated events may really originate from a statistically equilibrated source with $V=3.4 V_0$.}

Some considerations are now in order. All fragment properties
at the {\it dynamical freeze-out} appear to be almost univocally
linked to a statistical emission from a more compact equilibrated
source at an earlier time-step, {\it equilibrium freeze-out}, that
eventually expands on Coulomb trajectories. This time 
($t \simeq 140-150fm/c$) can be interpreted as when the dynamical 
evolution of the system reaches a stage of global equilibrium.

The {\it equilibrium freeze-out} time range appears just intermediate between the beginning of the 
spinodal decomposition and the final fragment configuration. At that time the
leading unstable modes are well established and some pre-fragments, in
strong interaction with the rest of the system, can be recognized. 
In this situation a statistical fragmentation
picture appears adequate, apart some minor effects, as the few
"equal fragment size" events, relics of the space structure of the primordial
spinodal instability [9].
However, due to the expansion of the system, the  {\it dynamical freeze-out}
configuration, where fragments are well formed and separated, corresponds
to a larger volume with respect to this {\it equilibrium freeze-out}. 


Since in dynamical simulations fragment formation is coupled with monopole oscillations of the source, so the system may recontract in some events or evolve towards more dilute configurations in some other cases \cite{buu} it is obvious
that we have to deal with the fluctuation of the freeze-out volume from event to event. On the other hand, in MMM calculations while $V/V_0$ is fixed, $\widetilde{V}/V_0$ is fluctuating. Whether or not the fluctuations of $\widetilde{V}/V_0$ are sufficient to mimic the ``dynamical'' ones or one should explicitly include fluctuations of $V/V_0$ it remains an open question. Since larger fluctuations are reflected into Coulomb effects it would be interesting
for future studies to compare the kinetic energy distributions around the average of Fig.4 (top) for each charge. 

From the nuclear thermodynamics point of view an important conclusion is that one deals with
{\em small} freeze-out volumes (3.4 $V_0$) and highly excited primary fragments (around 4.3 MeV/u), 
 i.e. {\em hot fragmentation} at {\em small} freezeout volumes. As it is well known, volume is a key variable 
 for locating the system in the phase diagram of the nuclear matter and a lot of work aiming to identify this quantity 
  is in progress \cite{borderie1}. 

Summarizing, along the fragmentation path, as described by a dynamical model, a huge part of the available phase space is filled. Statistical equilibration occurs in a pre-fragment stage of the system. To our knowledge, this is the first time when in dynamical paths of violent heavy ion  collisions a stage of statistical equilibrium of the compound system is proved. 

\begin{acknowledgments}
This work was partly supported by the European Community under a Marie Curie fellowship, contract n. MEIF - CT - 2005 - 010403  
\end{acknowledgments}

\end{document}